# Overcoming bias in representational similarity analysis

Roberto Viviani[1,2]

1: Institute of Psychology, University of Innsbruck, Austria
2: Psychiatry and Psychotherapy Clinic III, University of Ulm, Germany

Roberto Viviani
University of Innsbruck
Institute of Psychology
Innrain 52
6020 Innsbruck, Austria
roberto.viviani@uibk.ac.at

Abstract

Representational similarity analysis (RSA) is a multivariate technique to investigate cortical representations of objects or constructs. While avoiding ill-posed matrix inversions that plague multivariate approaches in the presence of many outcome variables, it suffers from the confounder arising from the non-orthogonality of the design matrix. Here, a partial correlation approach will be explored to adjust for this source of bias by partialling out this confounder in the context of the searchlight method for functional imaging datasets. A formal analysis will show the existence of a dependency of this confounder on the temporal correlation model of the sequential observations, motivating a data-driven approach that avoids the problem of misspecification of this model. However, where the autocorrelation locally diverges from its volume estimate, bias may be difficult to control for exactly, given the difficulties of estimating the precise form of the confounder at each voxel. Application to real data shows the effectiveness of the partial correlation approach, suggesting the impact of local bias to be minor. However, where the control for bias locally fails, possible spurious associations with the similarity matrix of the stimuli may emerge. This limitation may be intrinsic to RSA applied to non-orthogonal designs. The software implementing the approach is made publicly available (https://github.com/roberto-viviani/rsa-rsm.git).

# Overcoming bias in representational similarity analysis

## Introduction

Representational similarity analysis (RSA, Kriegeskorte et al. 2008) is a multivariate technique to investigate cortical representations of objects or constructs defined by a set of properties (the stimuli, see also Edelman et al. 1998). A defining characteristic of this approach is that it eschews inversion of estimates of large covariance matrices, an ill-posed problem that easily gives rise to unstable estimates. Instead, RSA seeks evidence of cortical encoding of properties of stimuli by assessing the concordance between the matrix encoding the pairwise dissimilarity of stimuli and a matrix encoding the dissimilarity of the brain responses to the same stimuli (in an equivalent alternative formulation, which will be followed here, the concordance is based on measures of similarity). Combined with the searchlight approach (Kreigeskorte et al. 2006), RSA allows mapping the cortex to seek evidence for the representation of stimuli based on their properties as defined by the experimenter, usually in the context of functional imaging experiments.

A complication of functional imaging is that brain responses are not measured directly but are mediated by the BOLD response. Hence, the brain signal to the presentation of a stimulus is estimated through the size of coefficients of a linear model encoding a canonical BOLD response. As it will be shown in detail below, a downside of RSA is that, unless the coefficients of this model were obtained with an orthogonal design, the covariance of these coefficients is confounded by the covariance arising from the design. Orthogonality is difficult to achieve because the sluggishness of the BOLD response introduces correlation between regressors of nearby trials. Because measures of similarity or dissimilarity between the brain signals are a function of their covariance, this confounder contaminates RSA estimates. Cai et al. (2019) have drawn attention to this problem, and proposed a maximum-likelihood/Bayesian approach to overcome it. As in another related approach (Friston et al. 2019), this requires knowing the form of the covariance of the coefficients to model it parametrically. In contrast with this literature, the present work addresses the case when it cannot be assumed that the form of the covariance is known. Instead, it explores the effectiveness of estimating this covariance directly from the data of the whole volume and adjust for it using partial correlation. This approach is reminiscent of PET techniques of adjusting model coefficients for global signal levels, except that the global term here is a covariance, and adjustment takes place via partial correlation.

This paper is structured as follows. In the theory section a formal treatment will identify the source of bias in RSA by tracing its connection with multivariate regression. It will first be shown that a simple RSA estimator of similarity of the brain signal is biased, also tracing the connection between RSA and multivariate regression. White noise data will then be used to demonstrate that bias can be positive or negative, depending on the relative position of labels. To show this, we will use a somewhat contrived design that represents a limit case in which bias in either direction can be produced. This limit case also allows to show why permutation approaches do not compensate for bias and may lead to false positives.

This formal treatment will then be the basis for the rationale of the partial correlation approach, also justifying its use in the face of other possible approaches, such as whitening. The same limit case design, where bias is particularly strong, will be used to test the capacity of the proposed approach to overcome bias in the white noise data as well as in EPI data. It will be shown that when adjusting for bias, a source of residual bias may arise from misidentifying the form of the autocorrelation of errors in the linear model, i.e. the lack of independence of sequential observations, as this model determines the covariance of the coefficients. We will explore the effectiveness of estimating this covariance from the volume data and apply the proposed approach data acquired at short TRs (where lack of independence of sequential observations may be pronounced).

Finally, we will present a case study of the application of the partial correlation approach to a more typical dataset where representation of body parts in the lateral occipital cortex may be expected based on the existing literature. As may expected from a design not intended to produce it, bias was minor in this dataset. Nevertheless, it led to significant clusters at testing, which disappeared after bias correction to be replaced by a single cluster, whose validity will be verified by comparing it to the meta-analytic location of the representation of body parts. The Discussion will mention the limitations of the RSA with the partial correlation approach, which are related to the difficulty of estimating the autocorrelation model accurately when it varies across the volume. The Material and Methods section will contain details on the acquisition parameters used in the MRI data.

The approach presented here was developed as an add-on function to the SPM package and is publicly available at https://github.com/roberto-viviani/rsa-rsm.git. At the present time, it is the only publicly available software to correct for RSA bias in the searchlight approach.

## Results

### Theory and simulations of RSA bias, ordinary least squares

When collected from a searchlight, the model may be written

$$\mathbf{Y} = \mathbf{XB} + \mathbf{E},$$

where $\mathbf{Y}$ is the $n$ scans $\times$ $p$ searchlight voxels matrix of the imaging data, $\mathbf{X}$ is a $n \times q$ design matrix usually containing $q$ BOLD-convolved regressors for the signal elicited by the presentation of stimuli during the experiment, $\mathbf{B}$ is a $q \times p$ matrix of coefficients to be estimated by least squares, and $\mathbf{E}$ the $n \times p$ matrix of errors. This is the multivariate regression model, which is applied repeatedly to each searchlight; to simplify notation, indexes referring to the searchlight are omitted. It will also be initially assumed that the rows of $\mathbf{E}$ be independent, an assumption that will be relaxed later. Well known results of multivariate regression are that the least squares estimates of $\mathbf{B}$ are identical to those obtained by regressing on the individual columns of $\mathbf{Y}$ separately (as in "the massively univariate approach" commonly adopted in functional imaging analyses), and that the expectation of these estimates and the expected variance-covariance are given by

$$E(\hat{\mathbf{B}}^V) = \mathbf{B}^V, \quad \mathrm{Var}(\hat{\mathbf{B}}^V) = \mathbf{\Sigma} \otimes (\mathbf{X}'\mathbf{X})^{-1}. \tag{1}$$

In this expression, $E(\cdot)$ denotes expectation, $(\cdot)^V$ is the operator that stacks the columns of a matrix upon each other into a vector, $\mathbf{\Sigma}$ is the $p \times p$ variance-covariance of the columns of the errors (i.e., over voxels), and $\otimes$ is the Kronecker product. From equation (1), the expected variance of a column $\hat{\boldsymbol{\beta}}_i$ of $\hat{\mathbf{B}}$ (the coefficient estimates from voxel $i$, $i = 1, \ldots p$) is

$$\mathrm{Var}(\hat{\boldsymbol{\beta}}_i) = \sigma_i^2 (\mathbf{X}'\mathbf{X})^{-1}, \tag{2}$$

where $\sigma_i^2$ is the diagonal term of $\mathbf{\Sigma}$ giving the variance of the errors in voxel $i$ (Mardia et al. 1979, pp. 160-161, 180-181).

After estimating the model coefficients $\hat{\boldsymbol{\beta}}_i$, RSA proceeds by selecting a set of coefficients or contrasts of coefficients for the analysis. Without loss of generality it will be assumed here that $\hat{\boldsymbol{\beta}}_i$ only refers to the required coefficients, usually obtained from a design in which the columns represent the expected BOLD response for each of the $q$ stimulus presentations. At this point, there are several possible approaches to compute the similarity or dissimilarity matrix of the brain signal, but they have similar effects and are all ultimately based on the inner product between the $q$ rows of $\hat{\mathbf{B}}$, taken in pairs (Kriegeskorte and Diedrichsen 2019). In all approaches, encoding of the stimuli in the searchlight is assessed by the correlation between the off-diagonal terms of a matrix collecting these measures with those of the analogous matrix of the stimuli.

Here, the matrix $\hat{\mathbf{B}}\hat{\mathbf{B}}' = \sum_i \hat{\boldsymbol{\beta}}_i \hat{\boldsymbol{\beta}}_i'$ (the sum of squares and cross products of the coefficients) will be used, as it is the simplest expression of these inner products and leads to the simplest derivations. Generalization to other approaches is motivated later. Note that $\hat{\mathbf{B}}\hat{\mathbf{B}}'$ is not a sum of squares and cross-products matrix in the usual sense, because rows are not independent (as specified by $(\mathbf{X}'\mathbf{X})^{-1}$ in eq. 1).

Starting from these properties of the multivariate regression model, it may be shown (see the Appendix) that

$$E(\hat{\mathbf{B}}\hat{\mathbf{B}}') = \mathbf{B}\mathbf{B}' + \sum_i \mathrm{Var}(\hat{\boldsymbol{\beta}}_i) \tag{3}$$

This result means that $\hat{\mathbf{B}}\hat{\mathbf{B}}'$ is a biased estimate of $\mathbf{B}\mathbf{B}'$, being confounded by $\sum_i \mathrm{Var}(\hat{\boldsymbol{\beta}}_i) = \sum_i \sigma_i^2 (\mathbf{X}'\mathbf{X})^{-1}$ (eq. 2 above; see also Cai et al. 2019 for a related expression). RSA bias arises from an accidental correlation between the off-diagonal terms of the similarity matrix of the stimuli and those of $(\mathbf{X}'\mathbf{X})^{-1}$ unless the columns of $\mathbf{X}$ are orthogonal and $(\mathbf{X}'\mathbf{X})^{-1} = \mathbf{I}$.

The crucial observation for the present approach is that the confounder due to $\sum_i \sigma_i^2 (\mathbf{X}'\mathbf{X})^{-1}$ is the same over all voxels and searchlights up to a multiplicative factor given by the voxel error variances $\sigma_i^2$. Because also the similarity matrix of the stimuli is the same over all searchlights, this gives rise to a *global bias* whose intensity is modulated by the voxel error variances but has otherwise the same origin over the whole volume. In contrast, $\mathbf{B}\mathbf{B}'$ will usually vary between searchlights, and its off-diagonal terms will be correlated with the similarity matrix of stimuli only in few sparse areas. This suggests that the average RSA correlation from the searchlights over the whole volume may provide a rough but simple diagnostic for the existence of global bias (below, we refer to this diagnostic as the 'average volume correlations').

The global bias arising from the non-orthogonality of the design is illustrated in the simulations of Figure 1. BOLD-convolved regressors for trials modeling exposure to stimuli for three seconds were formed, grouped into 6 blocks of 4 trials each (Figure 1A). Regularly spaced events were assigned to two class labels in an interleaved order or assigning the labels to homogenous blocks (Figure 1B; for details for the parameters used in the simulation, see the Methods section). The same stimuli were used in two patterns by arranging the sequence of labels differently. Pattern A comprised the stimuli of two different categories (in red and blue) alternating between trials within the blocks. In pattern B, the stimuli within the blocks belonged to the same category, in alternate blocks. The similarity matrices for these two patterns are in Figure 1C. While not representative of those found in typical RSA studies, this design was chosen because it causes a large amount of bias in different directions, giving an insight on the conditions that produce negative or positive bias. Figure 1D shows boxplots of the average volume correlations obtained by conducting a searchlight RSA of the similarity matrices of these two patterns after fitting the model to white noise data, where $\mathbf{BB'}$ is zero. One can see from the boxplots that pattern A resulted in artificially inflated positive correlations, while pattern B produced the opposite effect. Because the data are random noise, these correlations can only arise from $\sum_i \mathrm{Var}(\hat{\boldsymbol{\beta}}_i)$, which here is a multiple of $(\mathbf{X'X})^{-1}$, shown in panel A. One can see that the elements near the diagonal of this matrix follow pattern A, while pattern B groups together coefficients that are prevalently negatively correlated.

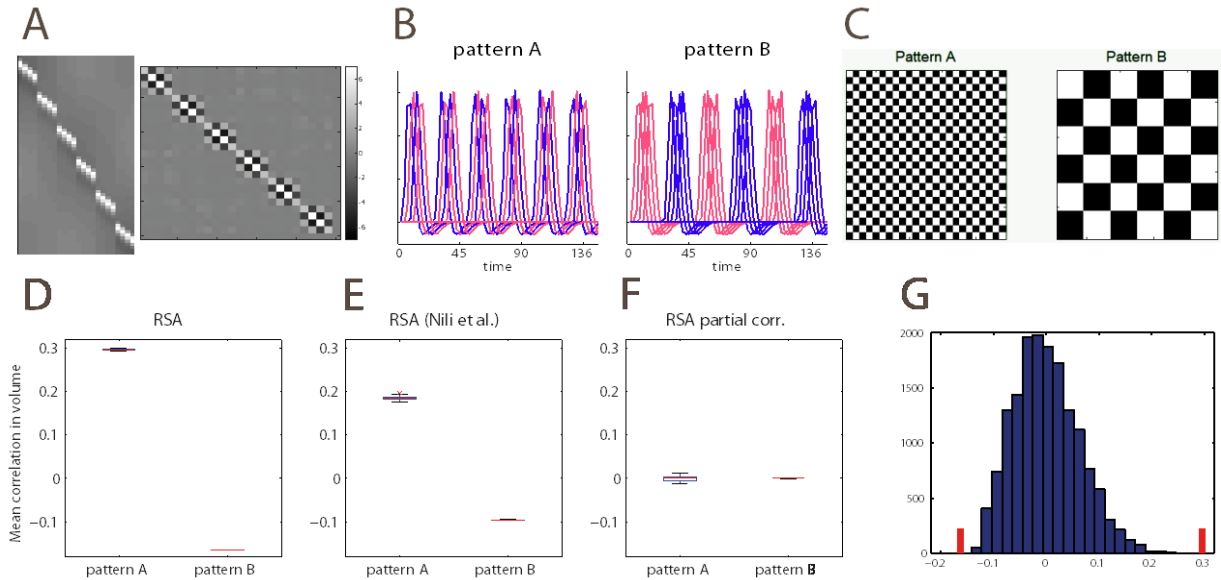

Figure 1. Simulation to elicit bias in RSA. A: Design matrix and covariance matrix (X'X)⁻¹ of the coefficients of the design. Covariances around zero are in gray, while black or white colors show negative and posiive covariances. B: classification of trials in two categories (shown in red and blue). This classification was chosen to produce correlation of coefficient models between trials of different categories ('pattern A') or between trials of the same category ('pattern B'). C: similarity matrices for patterns A and B. D and E: RSA for patterns A and B in volumes containing white noise, boxplots of estimated average volume correlations (y axis), showing values very different from zero. The results in panel E were computed with the original RSA toolbox (Nili et al. 2014). F: average volume correlations of RSA conducted with the partial correlation approach, after partialling out the effect of (X'X)⁻¹. One can see these correlations now being close to zero. G: Permutation distribution of randomizing category assignment and correlating the off-diagonal elements of the similarity matrices with (X'X)⁻¹. The red sticks on the x axis are the correlations obtained with patterns A and B, which are higher than any of the correlations obtained during the permutation.

Several approaches are conceivable when forming the similarity or dissimilarity matrix of the brain signal (Kriegeskorte et al. 2008), but they all involve inner products (Kriegeskorte and Diedrichsen 2019) collected in the matrix of squares and cross-products of the model coefficients. In the approach of Nili et al. (2014), the brain activation pattern is captured by a correlation matrix with an inverted sign, and the concordance with the stimulus pattern is assessed with rank correlation. Bias was apparent also with this approach (Figure 1E), computed with the original rsa toolbox (Nili et al. 2014). The bias amount differs because of the different approach in assessing the concordance between brain signal and stimuli. Relative to the sum of squares and cross products or a covariance matrix, correlation rescales the off-diagonal values, including those of $\sum_i \text{Var}(\hat{\boldsymbol{\beta}}_i)$, however without taking them to zero. Using ranks in the correlation affects the efficiency of the measured associations but does not abolish any accidental association with confounder patterns.

Using a permutation strategy by randomly reassigning the category labels of trials to compute significance (Kriegeskorte et al. 2008) will not help addressing the issue of bias, because the trials are not exchangeable. Here, patterns A and B represent cases in which categories are ordered so as to capture the patterns induced by the design matrix best, albeit in different ways, giving correlations in opposite directions. Most permutations will deliver label patterns that are some mixture of patterns A and B, resulting in global bias values intermediate between those obtained with these two patterns. Hence, the correlations due to bias obtained with patterns A and B will be located at the far tails of the permutation distribution, mistakenly suggesting their significance even if they were computed from noise (Figure 1G).

## Partial correlation approach to adjust for bias

As shown by the preceding analysis, bias arises from covariances in $\sum_i \text{Var}(\hat{\boldsymbol{\beta}}_i)$. A perhaps natural approach to make $\sum_i \text{Var}(\hat{\boldsymbol{\beta}}_i)$ diagonal would be to apply whitening. However, this approach will also represent the true similarity $\mathbf{BB}'$ in the modified outer product. Applying a whitening matrix $\mathbf{P}$ such that $\mathbf{P}^{\frac{1}{2}}\text{Var}(\boldsymbol{\beta}_i)\mathbf{P}^{\frac{1}{2}} = \sigma_i^2 \mathbf{I}$ to both sides of eq. (3) results in the replacement of $\mathbf{BB}'$ with $\mathbf{P}^{\frac{1}{2}}\mathbf{BB}'\mathbf{P}^{\frac{1}{2}}$:

$$\mathbf{P}^{\frac{1}{2}}\hat{\mathbf{B}}\hat{\mathbf{B}}'\mathbf{P}^{\frac{1}{2}} = \mathbf{P}^{\frac{1}{2}}[\mathbf{BB}' + \sum_i \text{Var}(\hat{\boldsymbol{\beta}}_i)]\mathbf{P}^{\frac{1}{2}}$$
$$= \mathbf{P}^{\frac{1}{2}}\mathbf{BB}'\mathbf{P}^{\frac{1}{2}} + \sum_i \sigma_i^2 \mathbf{I}$$

The off-diagonal values of the whitened sum of squares and cross-products are now an unbiased estimator, but of $\mathbf{P}^{\frac{1}{2}}\mathbf{BB}'\mathbf{P}^{\frac{1}{2}}$, not $\mathbf{BB}'$. For example, if $\mathbf{BB}' = p\mathbf{I}$ or some other multiple of the identity (i.e., if the coefficients of the model are independent and no similarity exists), then the off-diagonal terms are those of $\mathbf{P}$.

To remedy the effects of the design covariance of regression coefficients on RSA, the approach explored here consists in partialling out the effects of $\sum_i \text{Var}(\hat{\boldsymbol{\beta}}_i)$ on the off-diagonal terms of $\hat{\mathbf{B}}\hat{\mathbf{B}}'$ using partial correlation instead of using simple correlation to test for the presence of the similarity pattern in $\hat{\mathbf{B}}\hat{\mathbf{B}}'$. In the model of the simulations, this means using $(\mathbf{X}'\mathbf{X})^{-1}$ as a confounder. An informal justification for this approach is that, in a linear model, estimates of coefficients are unbiased irrespective of the amount of association of predictors with possible confounders. We should therefore be able to retrieve unbiased estimates of association between $\mathbf{BB}'$ and the similarity of stimuli irrespective of the accidental relation of these latter with $(\mathbf{X}'\mathbf{X})^{-1}$. As noted, the same may not be said for the whitening approach.

Figure 1F shows that the partial correlation approach succeeds in taking the average volume RSA correlations to zero in the white noise data. In this simple case, this is almost a given, because the confounder is exactly known up to the multiplicative factor of voxel error variance. This multiplicative factor is internally estimated by the partial correlation as the coefficient of the confounder term in each searchlight separately. However, if this term is not known exactly, residual confounding may ensue. In the following, the issue of modelling the confounder term is addressed for increasingly weaker assumptions.

## Generalized least squares

The results so far were obtained under the simplifying assumption that the observations were independent. Most functional imaging analysis approaches reject this assumption and fit a generalized least squares model, for example by including a first-order autoregressive term $\mathbf{G}$ estimated from residuals pooled over voxels,

$$\hat{\boldsymbol{\beta}}_i = (\mathbf{X}'\mathbf{G}^{-1}\mathbf{X})^{-1}\mathbf{X}'\mathbf{G}^{-1}\mathbf{y}_i.$$

The exact form of $\mathbf{G}$ depends on the model for the temporal autocorrelation of residuals and on implementation details (in the SPM package used in the present work $\mathbf{G}^{-1} = \mathbf{V}^{-\frac{1}{2}}\mathbf{H}_0\mathbf{V}^{-\frac{1}{2}}$, where $\mathbf{V}$ is an estimated AR(1) model for the residuals and $\mathbf{H}_0$ a residual-forming projection matrix implementing a high-pass filter). If this estimate is a good estimate of the true temporal dependency structure of the residuals, a standard result is that

$$\operatorname{Var}(\hat{\boldsymbol{\beta}}_i) \approx \sigma_i^2(\mathbf{X}'\mathbf{G}^{-1}\mathbf{X})^{-1}, \tag{4}$$

which replaces the corresponding term in eq. (3). Note that temporal whitening at the first level model is not sufficient to abolish the confounder arising from the design. On the contrary, now it is not certain that $\sum_i \operatorname{Var}(\hat{\boldsymbol{\beta}}_i)$ is diagonal even if the columns of the design matrix were orthogonal to each other (this will also be the case for ordinary least squares if the data are filtered). To adjust for the effect of the design, in datasets where generalized least squares is used one may now use the off-diagonal terms of Bcov = $(\mathbf{X}'\mathbf{G}^{-1}\mathbf{X})^{-1}$ to partial out its effects.

In practice, however, the autocorrelation model may not capture the true structure of the autocorrelation of the data. Furthermore, its estimate is usually computed from residuals pooled from multiple voxels, ignoring possible variations of the autocorrelation between voxels. Hence, a more realistic expression for the expected variance-covariance of the model coefficients may be given by

$$\operatorname{Var}(\hat{\boldsymbol{\beta}}_i) = \sigma_i^2(\mathbf{X}'\mathbf{G}^{-1}\mathbf{X})^{-1}\mathbf{X}'\mathbf{G}^{-1}\mathbf{\Gamma}_i\mathbf{G}^{-1}\mathbf{X}(\mathbf{X}'\mathbf{G}^{-1}\mathbf{X})^{-1} \tag{5}$$

(Diggle et al. 2002, p. 60), where $\mathbf{\Gamma}_i$ is the true dependency of the errors in voxel $i$. Only if $\mathbf{G}$ is a good estimate of $\mathbf{\Gamma}_i$ does this expression simplify to $\sigma_i^2(\mathbf{X}'\mathbf{G}^{-1}\mathbf{X})^{-1}$. This shows that, unless the autocorrelation term is appropriately modeled, residual confounding from the non-orthogonal design may still be present even after partialling out the effect of the off-diagonal terms of Bcov. Unfortunately, appropriately modeling error dependencies may be an arduous task because the true structure of $\mathbf{\Gamma}_i$ is unknown (i.e, whether AR(1) or something else) and may be influenced by study-specific factors (such as the properties of the MRI sequence, of the coil, and the frequencies of the regressors in the design matrix). As before, most terms of this expression are identical over the volume, inducing global bias, while the voxel-by-voxel variation of $\mathbf{\Gamma}_i$ introduces the possibility of *local bias*.

For the partial correlation approach, local bias means that one adjustment term alone may not control for the non-orthogonality of the design matrix satisfactorily. Nevertheless, even if $\mathbf{G}$ is not a good approximation of $\mathbf{\Gamma}_i$, it might be possible to address global bias by estimating the average variance-covariance $\hat{\text{Var}}(\mathbf{\beta})_*$ from the whole volume:

$$\hat{\text{Var}}(\mathbf{\beta})_* = \frac{1}{v} \sum_i (\hat{\mathbf{\beta}}_i - \bar{\mathbf{\beta}})(\hat{\mathbf{\beta}}_i - \bar{\mathbf{\beta}})', \qquad \bar{\mathbf{\beta}} = \frac{1}{v} \sum_i \hat{\mathbf{\beta}}_i . \qquad (6)$$

Here, the asterisk in the subscript indicates that the variance-covariance is estimated from the whole volume (as opposed to any individual searchlight), and $v$ is the number of voxels in the volume. One may then use the off-diagonal terms of $\mathsf{Svar} = \hat{\text{Var}}(\mathbf{\beta})_*$ in a partial correlation to control for global bias. The rationale for this approach is that global bias may be dominant given that most of the terms of eq. (5) are constant across the whole volume, and may be estimated as the average variance-covariance. A similar, even simpler approach consists of using the off-diagonal terms of $\mathsf{BB} = v^{-1}\hat{\mathbf{B}}_*\hat{\mathbf{B}}'_*$ to control for global bias, where the asterisk in the subscript of $\hat{\mathbf{B}}_*$ indicates again that this is the matrix of the coefficients for the whole volume. $\mathsf{BB}$ is confounded by the mean $\mathbf{BB}'_*$ over the whole volume (eq. 3). However, since the local $\mathbf{BB}'$ may be sparse and vary over searchlights, they may average out. With the due differences, use of $\mathsf{BB}$ as confounder is the closest equivalent of the 'global covariate' approach of PET analyses.

### Partial correlation in EPI data

Global bias and its correction with partial correlation RSA was explored in real MRI data using the same design as in Figure 1A (Labek et al. 2017, $N = 27$, TR = 2.27 sec; for details, see the Methods section), where bias correction is challenging because of its size. In Figure 2, panel A, global bias was assessed as in the previous simulations by looking at average volume correlations. There was evidence of global bias when computing the representational analysis without partialling out any confounder (ptn wo adjustment). This bias, obtained in real EPI data, was consistent with the findings of the simulations conducted on noise data, i.e. positive bias for pattern A and negative bias for pattern B. In panel B, one can see second-level $t$ maps of the RSA correlation for pattern A. The underlying average RSA volume correlations were positive in the whole volume, with peaks reaching $r = 0.15$, as one would expect in the presence of positive bias. These moderate correlations were sufficient to produce a $t$ map at the second level with values ranging between 8 and 25 across the volume. These $t$ values are high enough to cause rejection of the null hypothesis in all voxels in reasonable sample sizes.

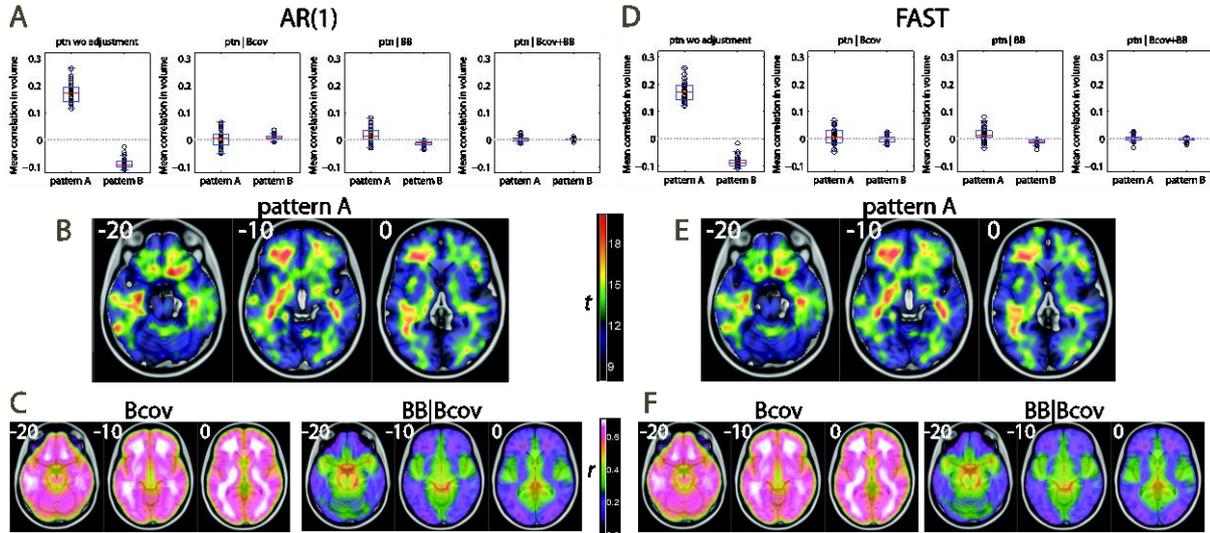

Figure 2. Analyses in real MRI data. Panels A-C: results obtained with an AR(1) for the lack of independence of sequential observations; panels D-F: same analyses obtained with FAST to model lack of independence. A and D: boxplots of mean volume correlations from the subjects in the sample. The boxplots on the left refer to RSA without partialling out confounder terms; from second left to right, to RSA when partialling out **Bcov**, **BB**, and both for patterns A and B (cf. Figure 1). B, E: *t* maps of RSA correlations without correction, showing large artefactual positive correlations for pattern A. C, F: searchlight analysis of the RSA correlation of the confounds and the data.

The parametric maps of the artefactual correlations of pattern A, shown in Figure 2B, were positive over the whole volume but of varying local intensity. Recall that the term 'global bias' refers to the identical source of bias in the whole volume given by the theoretical covariance of the model coefficients **Bcov** of eq. (4), but the intensity of this bias may be locally modulated by residual variance and the relative magnitude of **BB**′ in the individual searchlight. To assess its contribution, **Bcov** itself (as estimated from the AR(1) model of the pooled residuals by the SPM package) was used as a similarity map instead of pattern A in panel C, left. Like the spatial distribution of the artefactual effect of pattern A, the spatial distribution of the effect of **Bcov** prevalently affected the white matter region. Inspection of the artefactual correlation for pattern B revealed a similar effect over white matter, but in the negative direction (not shown here for brevity).

The effectiveness of the partial correlation adjustments is displayed in the remaining boxplots of panel A. Partialling out the off-diagonal terms of **Bcov** removed a large part of global bias (panel A, ptn|Bcov, second from left), as did using **BB** (ptn|BB, third from left), as shown by the average volume correlations being close to zero. However, only using both **Bcov** and **BB** simultaneously achieved results comparable to those of the simulations with noise (ptn|Bcov+BB, right), suggesting the existence of local bias (i.e., bias that could not be captured by one covariance term alone). Testing this latter RSA model using a permutation method to correct for multiple testing on leave-one-out resampling of the original data gave a false positive rate of 0.07 for $p < 0.05$.

To show where **BB** might have improved over **Bcov** in the combination, Panel C of Figure 2, right, displays the RSA of **BB** itself used as a similarity map after partialling out the effect of **Bcov**. The strongest correlations followed the distribution of vessel density (Viviani 2016), a main source of high variance in EPI data originating in blood pulsation (Lund et al. 2006). The increased signal on the midline is similar to effects of respiratory volumes reported by Birn et al. (2006). These voxels were missing from the RSA of **Bcov**, notwithstanding the high levels of signal noise, as one would expect if **Bcov** was less good as a model of the covariance of the coefficients in these voxels. There was no qualitative difference when using Svar instead of BB in these analyses.

In panels D-F of Figure 2 we repeated the same analyses using FAST (Corbin et al. 2018), an option in the SPM package to model autocorrelations more flexibly by including several exponential time terms and their derivatives. The results were similar to those of the AR(1) model, suggesting the simpler model to be adequate in this case.

A question raised by these data is whether the effectiveness of adjusting by Bcov may be due to the ease with which the autocorrelation of residuals is modelled in these data. In Figure 3 we applied the same partial RSA strategy to correct for global bias in MRI data collected for the present study with TR = 0.34 sec ($N = 32$), where the correlation of sequential acquisitions may be larger. Perhaps surprisingly, the extent of global bias was smaller here than in the previous sample. Nevertheless, there were widespread artefactual RSA correlations (panel B), following the spatial pattern of Bcov, which loaded above all on the center of the volume (where noise increases with distance from the coil) and to a lower extent on white matter (panel C). These findings are consistent with a much larger relative contribution of thermal noise in the variance of these data. The relative contribution of BB involved here gray matter. The control for global bias through partial correlation was equally effective as in the long TR data (panel A).

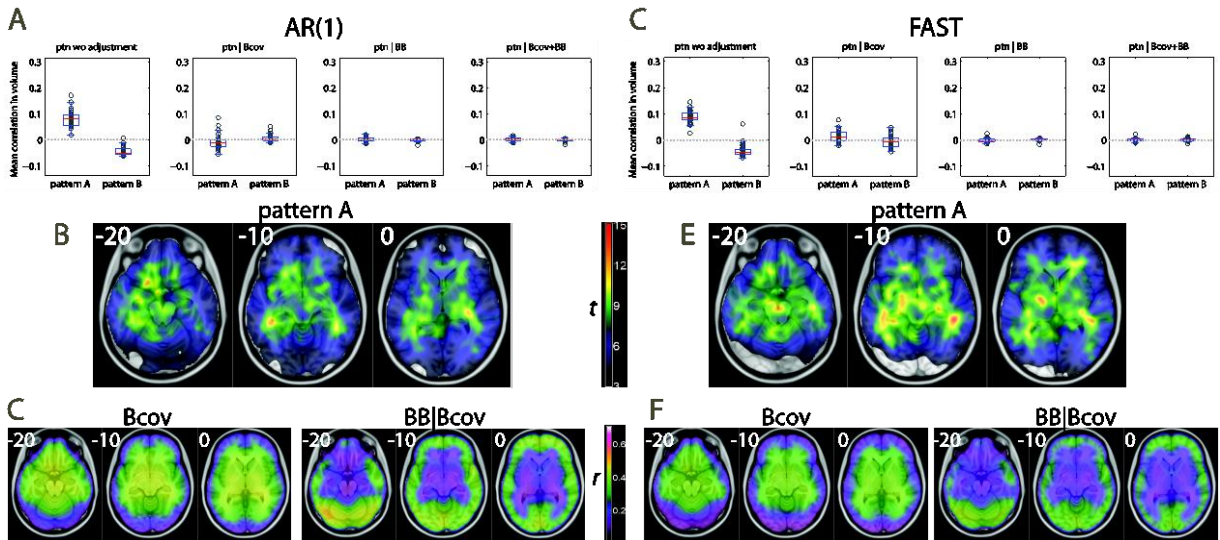

Figure 3. Analyses in real MRI data at short TR (same organization of panels as in Figure 2). Note that the scale of the t maps in B and E differs from the one used in Figure 2.

The use of FAST in this dataset (panels D-F) revealed the relative contribution of BB to be lower after adjusting for Bcov, as one would expect in a more realistic autocorrelation model of the error. However, we also observed high variation in the estimates of the autocorrelation term. There were no false positives in the leave-one-out resampling estimate of the FAST model after partialling out BB and Bcov at the $p < 0.05$ threshold, corrected at voxel level with the permutation method.

### Additional strategies for the control of bias

Several additional strategies to control for bias are conceivable in individual studies. One is limiting the volume mask to gray matter to make the residual correlation model more representative of the target brain areas. A more aggressive strategy consists in computing correlations between off-diagonal terms further away from the main diagonal, as those close to it, if the regressors for the single coefficients were arranged in temporal order in the design matrix, are those where the effect of the design is strongest. This means omitting the cross-products of trials that are temporally close to each other in the statistics, assuming that similarity can be estimated from further trials. This strategy may be thought of as a resampling alternative to whitening: while this latter weights the coefficients of nearby trials to adjust for the estimated covariance, here these trials are simply omitted, obviating the need for an accurate estimate of this covariance. If stimulus presentations are distributed equally in nearby and in far trials, the estimates of concordance may be unbiased by the same reasoning that applies to estimates from 'missing at random' data. In our dataset, implementing this strategy improved uncorrected global bias (Figure 4). However, partial correlation still appeared to improve the level of global bias in individual volumes in these data.

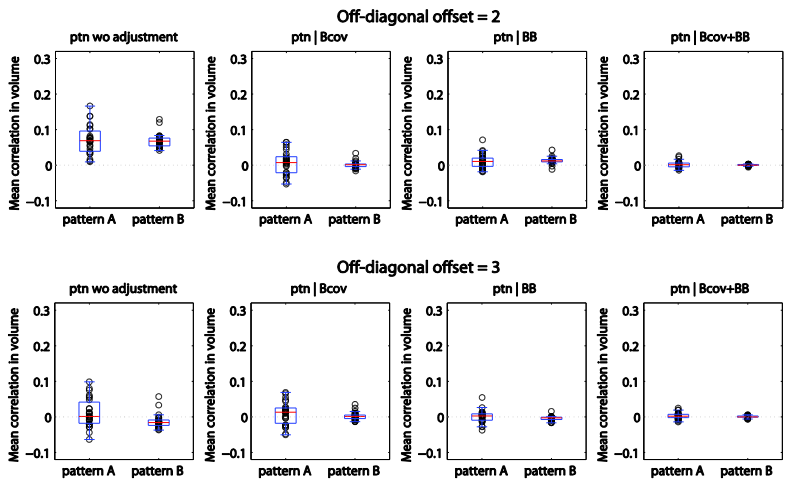

Figure 4. Average volume correlations estimates of global bias in the AR(1) model of long TR data when the sum of squares and cross products matrix is used with an off-diagonal offset of 2 and 3 units.

### Application to real pattern in EPI data

To illustrate the partial correlation approach, we analyzed the data of Figure 2, this time using the real pattern of the stimuli of the study, which were arranged in two categories (mourning individuals and control images) in a block pattern (like pattern B but with the correct category labels, Labek et al. 2017). A progressively more aggressive strategy was followed to adjust for bias, starting with Bcov alone and then combining Bcov with Scov and then BB, giving mean average volume correlations of 0.0053±0.0074, 0.0039±0.0044, and 0.0014±0.0026. Figure 5 shows the overlap between the contrast of interest (mourning vs. control, in light blue) and the parametric map of the one-sample $t$ test of the correlations obtained with partial correlation (blue-red-yellow colors). The cluster in the precuneus was significant when tested with a permutation approach in all adjustment models, but the Figure shows overlap also in clusters that did not survive correction.

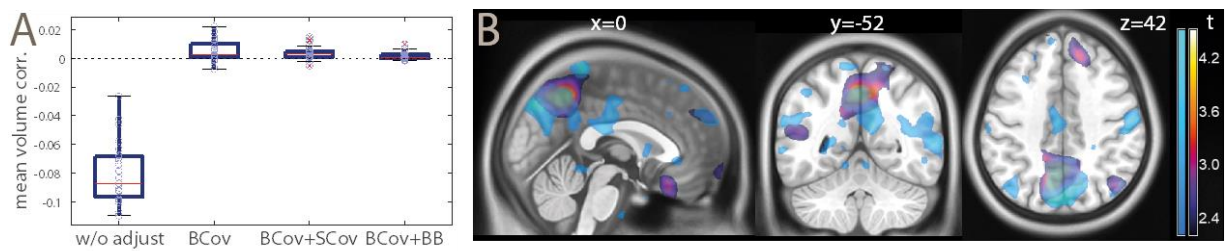

Figure 5. Overlap between the standard contrast mourning vs. control (in light blue) and the RSA analysis of the equivalent pattern (in blue-red-yellow) obtained with the adjustment terms BCov+SCov. For visualization purposes, parametric maps were thresholded at $p < 0.005$, uncorrected.

In contrast, when computing the RSA with a simple correlation, the accidental negative association between the blocks and the design covariance gave widespread negative correlations in the RSA analysis, more intense in white matter. Here, the mean average volume correlation was -0.08±0.021. While this bias may seem minor, the *t* values of the parametric map at the second level ranged between -3.3 and -38.3 across the volume, leading to rejecting the null in the vast majority of voxels in the negative direction. A symptom of this strong bias was that the standard deviation of the average volume correlations (0.021) was of smaller absolute magnitude than the mean (-0.08).

## Application to RSA of body parts

To illustrate the effectiveness of the partial correlation approach in adjusting for bias in a more typical RSA application, we analysed an experiment (Labek et al. 2021) in which participants were presented with two scenes depicting either a right hand or a right foot (these scenes were first used in Jackson et al. 2005). These scenes depict either an imminent injury or a neutral situation. More details on this experiment, including a full analysis of the cortical representation of pain, are presented in Labek et al. (2021). Here, we focus on the bias in the RSA of the body parts displayed in the trials (two body parts per trial, hands or feet) as a case study.

Even if the trials were separated by 10 sec or more, the RSA without bias adjustment showed a small negative bias (Figure 6A, w/o adjustment). This bias may be attributed to a correlation of the similarity matrix encoding body parts and the covariance due to the design, which was 0.04 for Bcov (this value included the estimated autocorrelation from pooled residuals, as computed by the SPM12 package). This negative bias was large enough to produce several negative correlation clusters, and no significant positive correlation (Figure 6B).

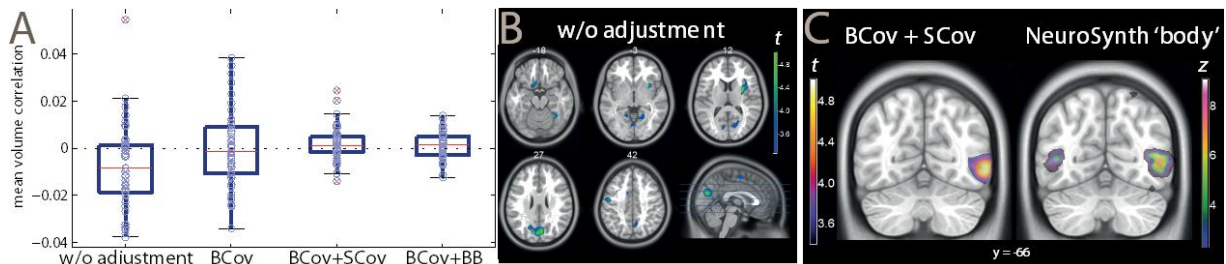

Figure 6. A: average volume correlations estimates of global bias in the RSA analysis for body parts without adjustment (left) and applying progressively more aggressive adjustment terms. B: significant negative correlation clusters in the RSA analysis without adjustment (cluster definition threshold, p < 0.001) from permutation testing. C: RSA analysis with the Bcov + Scov adjustment terms, producing one significant positive cluster and no significant negative clusters (left), and comparison with the NeuroSynth meta-analytic map for the term 'body' (right).

Adding Bcov alone or combining Bcov with Scov and then with BB was effective against bias, with Bcov adjustment alone already showing no evidence of bias (Figure 6A). All these adjustment models gave very similar results: no significant negative correlation in the RSA and a significant cluster of positive correlation in the right lateral occipital complex (significant at peak level correction *p* = 0.019-0.001; cluster level, *p* = 0.044-0.006; Figure 6C). While all the adjusted models gave similar RSA maps with the same cluster, there was no relation with the significant clusters of the unadjusted model. To validate the cluster of the adjusted analyses we used the meta-analytic tool NeuroSynth (Yarkoni et al. 2011). Comparison with the NeuroSynth images for the topics term 'body' or 'hand' revealed an almost perfect correspondence (Figure 6C).

**Discussion**

This work sought to address the confounder arising in RSA from the non-orthogonality of the design matrix using a partial correlation approach. Unfortunately, the precise form taken by the ensuing confounder also depends on the temporal autocorrelation of the residuals in successive MR acquisitions, a quantity that is difficult to estimate. The importance of this issue in RSA contrasts with its benign consequences in ordinary analyses of functional imaging experiments. Although the imaging literature contains numerous statements on the consequences of inadequate models of residual temporal correlation on correct significance levels, starting at least from Purdon and Weisskoff (1998), these consequences only concern inference at the first level (generalizing over scans within subjects), which is almost never considered in practice. Irrespective of residual correlations and how they are modelled, coefficient estimates (including OLS estimates) are always unbiased (Diggle et al. 2002), and in ordinary functional imaging studies it is these coefficient maps that are brought to the second level (to generalize over subjects), where they are independent. Hence, models of temporal autocorrelation in the 'massively univariate' approach of ordinary functional imaging studies only affect the efficiency of model estimates, not the correctness of inference at the second level. This fortunate state of affairs does not hold in RSA.

When the theoretical estimate of the covariance is adequate, its use as a confounder term is straightforward. This is a pragmatic approach, but it has a justification in the unbiasedness of coefficients estimates in a linear model even when there is substantial association with confounders. When residual temporal autocorrelation is not adequately modelled, as appeared to be the case in our data, we explored the effectiveness of correcting for the ensuing global bias by estimating the covariance of the coefficients of the first-level model from the volume data and partialling it out. It is straightforward to increase the effectiveness of confounder adjustment by adding different confounder terms in the model and allow their effect to vary over the volume. These advantages are difficult to obtain with other strategies.

The use of covariance estimates of coefficients from volume data assumes that the association of the off-diagonal terms of $\mathbf{BB}'$ and the stimuli similarity maps is present in few voxels, otherwise this association will be removed by the adjustment (Andersson et al. 2001). As in other global covariate approaches, if $\mathbf{BB}'$ has the same form in large parts of the volume, the adjustment might introduce some bias of its own in place of the bias due to the design (Aguirre et al. 1998). This situation is essentially the same that confronts PET data analyses. It is even more difficult to provide a comprehensive strategy to counter local bias. The results presented here suggest that local bias was minor in our data, as shown by the large effects of corrections obtained with single terms, but it would be difficult to generalize to all possible datasets. Alternatively, the off-diagonal offset approach aims at obtaining estimates of concordance irrespective of what the covariance between nearby trials is by exploiting a weaker assumption of equal distribution of category labels in nearby and far trials. The price to pay is the exclusion of some trials from the analysis, reducing sensitivity.

It might be appropriate to compare the effect of different combinations of correction strategies and error correlation models, as true correlations with the similarity matrix of the stimuli should be robust to these variations. However, the possible existence of local bias implies no guarantee that the set of correction terms used be equally and uniformly effective in all voxels. In this case, possible spurious associations with the similarity matrix of the stimuli may emerge. This is an issue that may be difficult to address unless efficient methods to model the local model covariance of coefficients are developed.

There are other limitation issues with the present work. One is that the limit case design used here is not representative of designs used in RSA studies. This argument is misguided, since one single counterexample suffices to demonstrate the failure of an approach. Once the problem is exposed, the onus of showing lack of bias rests on individual experimenters, not on us showing how frequent bias is in the existing literature. The design explored here was chosen because it considers the limit cases for positive and negative bias in the analysis, clarifying why bias may occur, not because it is representative of designs used in practice. Furthermore, we have shown that bias may occur in trials separated by 10 sec, demonstrating that it is a concrete eventuality.

It is also important to underscore the fact that experimenters have recourse to planning the design so as to minimize the effects of the confounder arising from the association from the model coefficients. It may also be possible to counteract bias by computing the representational similarity analysis on chosen contrasts grouping trials so as to alter the extent of the association between the contrast coefficients, instead of using a design in which each trial is modelled by a single regressor. Furthermore, "condition-rich" designs, which are those with many features and thought best suited for representational similarity analysis (Kriegeskorte et al. 2008) may present less marked accidental associations with the pattern induced by the design. However, in the absence on corroborating information, this remains only a possibility. Furthermore, application of RSA to the study of abstract features, such as those of high level cognition (Freund et al. 2021) might require modelling low-dimensional categories, where the examples shown here may be considered typical.

## Methods

The simulations of Figure 1 were conducted in MATLAB. White noise volumes of isotropic voxel size 2mm were used as input data, creating 65 volumes for each of $N = 30$ subjects. The block design was based on Labek et al. (2017).

This design was also used in the data of the first two datasets. Participants viewed drawings of individuals, each displayed for 3 sec. in blocks of 4 drawings. Hence, each block had a duration of 12 sec. Scrambled drawings were displayed between blocks for the same duration to form a high-level baseline. There were three blocks of drawings depicting sad individuals and three blocks of control drawings. Drawings and blocks were assigned to arbitrary stimuli classes to form the patterns of Figure 1. Both EPI samples were acquired on the premises of the Clinic for Psychiatry and Psychotherapy of the University of Ulm, Germany, after obtaining written informed consent from volunteers recruited locally. The study was approved by the Ethical Committee of the University of Ulm.

In the analysis of Figure 2, data were collected with a T2*-weighted EPI in a Siemens 3T Allegra scanner (TR/TE 2260/35 msec, flip angle 90°, in-planar voxel size 3x3mm, slice thickness 3mm, 25% gap between slices, 35 slices) with a single-channel head coil. In each of $N = 27$ participants, 65 volumes were acquired. The data of Figure 3 were acquired in a Siemens 3T Prisma scanner with an EPI multiband sequence, with parameters chosen to minimize TR (slice acceleration factor 6, TR/TE 340/50 msec, flip angle 90°, in-planar voxel size 3x3mm, slice thickness 3mm, 13% gap between slices, 32 slices). No attempt was made to optimize for BOLD sensitivity, but we verified that the task and the contrast reported by Labek et al. (2017) were present in these data. Different pictures were used here for purposes extraneous to the present study. In each of $N = 32$ participants, 433 images were acquired. To assist registration, a short T2-weighted EPI sequence of 5 volumes was acquired with no acceleration (TR/TE 4500/13 msec, flip angle 90°, same geometry as in the accelerated sequence). A 64-channels head coil was used with foam padding to minimize head motions.

The body parts dataset was acquired in a decision-making study (Labek et al. 2021). In each trial, two scenes were presented side by side and participants were asked to press a button on the side of the more painful situation. All combinations of scenes were presented (i.e. painful and neutral, neutral and neutral, and painful and painful). Trials lasted between 2.5 and 4 sec, and occurred at intervals of about 14.75 sec on average. More details on this experiment, including a full analysis of the cortical representation of pain, are presented in Labek et al. (2021). Data were collected using a $T2^*$-sensitive echo-planar imaging sequence (TR/TE 1970/36 msec, bandwidth 1776 Hz/pixel, slice thickness 2.5 mm with a slice gap of 0.625 mm, field of view 192 with matrix size 64, giving in-planar voxel size 3x3 mm, slice thickness 2.5 mm with a slice gap of 0.625 mm,) in a Prisma 3T scanner (Siemens, Erlangen) with a 24-channels head coil. In each participant, 182 scans were acquired after reaching equilibration giving a total scan duration of 6 min.

The EPI data were preprocessed with SPM12 (Functional Imaging Laboratory, London), using standard procedures (realignment, normalization), producing data resampled at the isotropic voxel size 2mm. Normalization of the multiband data was based on multimodal segmentation, which included the mean multiband and the mean EPI without acceleration volumes simultaneously. The success of the segmentation was verified by inspection. Coefficients of noise and EPI data were modelled at the first level by convolving a stick function of duration 3 sec with a canonical BOLD curve. Each drawing presentation was modelled by its own regressor, giving as many beta images as stimuli (as suggested by Kriegeskorte et al. 2008). An intercept and realignment parameters were added as confounding covariates. In the analysis of Figure 5, for denoising four confounding covariates were added from the first four principal components from tissue classified as lateral ventricles, white matter, or bone in the segmentation internally used by SPM in the normalization step. No high-pass filter, autocorrelation model, or confounding covariates were applied to noise data. The high-pass filter of EPI data was set at 128 sec. No smoothing was applied to volume data prior to computing the first level model.

The RSA was coded in MATLAB as an extension of the SPM package using the `spm_searchlight` function offered in that package. The spherical searchlight had a dimension of 8 mm, including 257 voxels when not located near the edge of the selected volume, defined by a mask including grey and white matter as well as CSF using the prior provided by SPM for these compartments at threshold $p < 0.5$. Searchlight volumes including less than 27 voxels were excluded. The similarity matrix contained the value zero for entries of different categories, and a constant value for entries of the same

– 15 –

categories. As described in the main text, the similarity of the brain response was encoded as the sum of squares and cross products of the coefficients of the design shown in Figure 1. Pearson correlations and partial correlations between the upper diagonal terms of these two matrices were computed using the native MATLAB functions. Bcov was extracted from the internal SPM computation of that quantity, whereas Scov and BB were computed as indicated by the equations in the main text. The results in Figure 1C were computed with the function `searchlightMapping_fMRI` from the toolbox described by Nili et al. (2014), which uses Spearman correlation to assess correspondence.

The overlays of panels B and E of Figures 2 and 3 were obtained by computing $t$ maps of one-sample $t$ test of the searchlight correlations brought to the second level and smoothed with a Gaussian kernel (FWHM 4mm). The overlays of panels C and F are average maps of searchlight correlations (same smoothing). The false positive rates reported in text were computed by resampling the data without replacement, each time leaving out one individual at the second level, and computing significance levels with a permutation method for strong control of significance levels (percentiles of maximal $t$ values over the volume, 2000 permutations per resampling; see Holmes et al. 1996). The rates are the percent of resamples where the null was rejected at the voxel level-corrected significance threshold $p < 0.05$.

## Appendix

Recall that the least squares estimates of coefficients in the multivariate regression model $\mathbf{Y} = \mathbf{XB} + \mathbf{E}$ are identical to those obtained by regressing on the individual columns of $\mathbf{Y}$ separately (see the main text for notation). Considering the coefficients $\hat{\boldsymbol{\beta}}_i$ of the regressions on the individual columns of $\mathbf{Y}$, a standard result is the following,

$$\hat{\boldsymbol{\beta}}_i = \boldsymbol{\beta}_i + (\mathbf{X}'\mathbf{X})^{-1}\mathbf{X}'\boldsymbol{\varepsilon}_i$$
$$= \boldsymbol{\beta}_i + (\hat{\boldsymbol{\beta}}_i - \boldsymbol{\beta}_i)$$

where $\boldsymbol{\varepsilon}_i$ is a column of $\mathbf{E}$ (Mardia et al. 1979, pp. 160). Taking the squares of both sides of this equation and summing over the voxels indexed by $i$, one obtains

$$\sum_i \hat{\boldsymbol{\beta}}_i \hat{\boldsymbol{\beta}}_i' = \sum_i \boldsymbol{\beta}_i \boldsymbol{\beta}_i' + \sum_i (\hat{\boldsymbol{\beta}}_i - \boldsymbol{\beta}_i)(\hat{\boldsymbol{\beta}}_i - \boldsymbol{\beta}_i)' + \sum_i \boldsymbol{\beta}_i (\hat{\boldsymbol{\beta}}_i - \boldsymbol{\beta}_i)' + \sum_i (\hat{\boldsymbol{\beta}}_i - \boldsymbol{\beta}_i) \boldsymbol{\beta}_i'.$$

When taking expectations, the cross-product terms $\sum_i \boldsymbol{\beta}_i (\hat{\boldsymbol{\beta}}_i - \boldsymbol{\beta}_i)' + \sum_i (\hat{\boldsymbol{\beta}}_i - \boldsymbol{\beta}_i) \boldsymbol{\beta}_i'$ vanish, giving

$$E(\hat{\mathbf{B}}\hat{\mathbf{B}}') = \mathbf{B}\mathbf{B}' + \sum_i \mathrm{Var}(\hat{\boldsymbol{\beta}}_i).$$

Replacing $(\mathbf{X}'\mathbf{X})^{-1}\mathbf{X}'$ above with the corresponding linear operator of GLS regression generalizes this proof to eqs. (4) and (5) of the main text.

## Acknowledgments

Permission to use images from previous work by Jean Decety is gratefully acknowledged. This work was conducted within the framework of the "Austrian NeuroCloud", supported by the Austrian Federal Ministry of Education, Science and Research. Acquisition of data was supported by a collaborative grant from the Federal Institute for Drugs and Medical Devices (BfArM, Bonn, Grant No. V-17568/68502/2017-2020). The author declares no conflict of interest.

**References**

Aguirre, G.K., Zarahn, E., D'Esposito, M., 1998. The inferential impact of global signal covariates in functional neuroimaging analysis. NeuroImage 8, 302-306.

Andersson, J.L., Ashburner, J., Friston, K., 2001. A global estimator unbiased by local changes. NeuroImage 13, 1193-1296.

Birn, R.M., Diamond, J.B., Smith, M.A., Bandettini, P.A., 2006. Separating respiratory-variation-related fluctuations from neuronal-activity-related fluctuations in fMRI. NeuroImage 31, 1536-1548.

Cai, M.B., Schuck, N.W., Pillow, J.W., Niv, Y., 2019. Representational structure of task structure? Bias in neural representational similarity analysis and a Byesian method for reducing bias. PLoS Comp. Biol. 15(5), e1006299.

Corbin, N., Todd, N., Friston, K.J., Callaghan, M.F., 2018. Accurate modeling of temporal correlations in rapidly sampled fMRI time series. Hum. Br. Mapping 39, 3882-3897.

Diggle, P.J., Heagerty, P., Liang, K.Y., Zeger, S.L., 2002. Analysis of Longitudinal Data (2nd ed.). Oxford University Press, Oxford.

Edelman, S., Grill-Spector, K., Kushnir, T., Malach, R., 1998. Toward direct visualization of the internal shape representation space by fMRI. Psychobiol. 26, 309-321.

Freund, M.C., Etzel, J.A., Braver, T.S., 2021. Neural coding of cognitive control: The representational similarity analysis approach. Trends Cogn. Sci. (in press), .

Friston, K.J., Diedrichsen, J., Holmes, E., Zeidman, P., 2019. Variational representational similarity analysis. NeuroImage 201, 115986.

Holmes, A.P., Blair, R.C., Watson, J.D.G., Ford, I., 1996. Nonparametric analysis of statistic images from functional mapping experiments. J. Cereb. Blood Flow Metab. 16, 7-22.

Jackson, P.L., Meltzoff, A.N., Decety, J., 2005. How do we perceive the pain of others? A window into the neural processes involved in empathy. NeuroImage 24, 771-779.

Kriegeskorte, N., Diedrichsen, J., 2019. Peeling the onion of brain representations. Annu. Rev. Neurosci. 42, 407-432.

Kriegeskorte, N., Goebel, R., Bandettini, P., 2006. Information-based functional brain mapping. Proc. Natl Acad. Sci. USA 103, 3863.3868.

Kriegeskorte, N., Mur, M., Bandettini, P., 2008. Representational similarity analysis. Connecting the branches of systems neuroscience. Frontiers Sys. Neurosci. doi: 10.3389/neuro.06.004.2008.

Labek, K., Berger, S., Buchheim, A., Bosch, J., Spohrs, J., Dommes, L., Beschoner, P., Stingl, J.C., Viviani, R., 2017. The iconography of mourning and its neural correlates: A functional neuroimaging study. Soc. Cogn. Aff. Neurosci. 12, 1303-1313.

Labek, K., Sittenberger, E., Kienhöfer, V., Rabl, L., Messina, L., Schurz, M., Stingl, J.C., Viviani, R., 2021. The gradient model of brain organization in decisions involving 'empathy for pain'. bioRxiv doi: https://doi.org/10.1101/2021.11.28.470235.

Lund, T.E., Madsen, K.H., Sidaros, K., Luo, W.-L., Nichols, T.E., 2006. Non-white noise in fMRI: Does modelling have an impact? NeuroImage 29, 54-66.

Mardia, K.V., Kent, J.T., Bibby, J.M., 1979. Multivariate Analysis. Academic Press, London.

Nili, H., Wingfield, C., Walther, A., Su, L., Marslen-Wilson, W., Kriegeskorte, N., 2014. A toolbox for representational similarity analysis. PLoS Comp. Biol. 10, e1003553. Software available at https://www.mrc-cbu.cam.ac.uk/methods-and-resources/toolboxes/ (last visited 16 Sept. 2020).

Purdon, P.L., Weisskoff, R.M., 1998. Effect of temporal autocorrelation due to physiological noise and stimulus paradigm on voxel-level false-positive rates in fMRI. Hum. Br. Mapping 6, 239-249.

Viviani, R., 2016. A digital atlas of middle to large brain vessels and their relation to cortical and subcortical structures. Frontiers Neuroanat. 10, 12.

Yarkoni, T., Poldrack, R.A., Nichols, T.N., Van Essen, D.C., Wager, T.D., 2011. Large-scale automated synthesis of human functional neuroimaging data. Nat. Meth. 8, 665-670.